\documentclass[11pt,moriond]{article}
\usepackage{moriond,epsfig}

\def\Journal#1#2#3#4{{#1} {\bf #2}, #3 (#4)}

\usepackage{natbib}

\def\NIM{\em Nucl. Instrum. Methods}
\def\NIMA{{\em Nucl. Instrum. Methods} A}

\def\be{\begin{equation}}
\def\ee{\end{equation}}
\def\bea{\begin{eqnarray}}
\def\eea{\end{eqnarray}}

\begin{document}
\vspace*{4cm}
\title{DISCOVERY OF VHE $\gamma$-RAYS FROM CENTAURUS A}

\author{M. RAUE$^1$, J.-P. LENAIN$^3$, F. A. AHARONIAN$^{7,1}$, Y. BECHERINI$^2$, C. BOISSON$^3$, A.-C.~CLAPSON$^1$, L. COSTAMANTE$^6$, L. G\'ERARD$^2$, C. MEDINA$^3$, M.~DE NAUROIS$^4$, M.~PUNCH$^2$, F.~RIEGER$^1$, H.~SOL$^3$, \L. STAWARZ$^5$, A. ZECH$^3$, for the H.E.S.S. COLLABORATION}

\address{$^1$Max-Planck-Institut fuer Kernphysik, Heidelberg, Germany,
$^2$Astroparticule et Cosmologie (APC), CNRS, Universit\'e Paris 7 Denis Diderot,
France,
$^3$LUTH, Observatoire de Paris, CNRS, Universit\'e Paris Diderot, Meudon, France,
$^4$LPNHE, Universit\'e Pierre et Marie Curie Paris 6, Universit\'e Denis Diderot Paris 7, CNRS/IN2P3, France,
$^5$Astronomical Observatory, Jagiellonian University, Cracow, Poland,
$^6$HEPL/KIPAC, Stanford University, US,
$^7$Institute for Advanced Studies, Dublin, Ireland}

\maketitle\abstracts{
We report the discovery of faint very high energy (VHE, $E > 100$\,GeV) $\gamma$-ray emission from the radio galaxy Centaurus A in deep observations performed with the H.E.S.S. experiment. A signal with a statistical significance of 5.0\,$\sigma$ is detected from the region including the radio core and the inner kpc jets. The integral flux above an energy threshold of $\sim$250\,GeV is measured to be $~0.8\,\%$ of the flux of the Crab Nebula and the spectrum can be described by a power law with a photon index of $2.7 \pm 0.5_{\mathrm{stat}} \pm 0.2_{\mathrm{sys}}$. No significant flux variability is detected in the data set. The discovery of VHE $\gamma$-ray emission from Centaurus A reveals particle acceleration in the source to $>$\,TeV energies and, together with M\,87, establishes radio galaxies as a class of VHE emitters.}


\section{Introduction}

Centaurus~A (Cen~A) is the nearest active radio galaxy (3.8\,Mpc \cite{rejkuba:2004a}; for a review see \cite{israel:1998a}).  At radio wavelengths rich jet structures are visible, extending from the core and the inner pc and kpc jet to giant outer lobes with an angular extension of $8^{\circ} \times 4^{\circ}$. The inner kpc jet has also been detected in X-rays, revealing a complex structure of bright knots and diffuse emission \cite{kraft:2002a}. The angle of the jet axis to the line of sight is estimated to be 15-80$^\circ$ \cite[see e.g.][and references therein]{horiuchi:2006a}. The elliptical host NGC~5128 features a dark lane, a thin edge-on disk of dust and young stars, believed to be the remnant of a merger. Recent estimates for the mass of the central supermassive black hole give $ (5.5 \pm 3.0)  \times 10^{7}$\,$M_\odot$ \cite{cappellari:2008a}.
The kpc-scale jet and the active nucleus are confirmed sources of strong non-thermal emission. In addition, more than 200 X-ray point sources with an integrated luminosity of L$_{\mathrm{X}} > 10^{38}$\,erg\,s$^{-1}$ are established to be associated with the host galaxy \cite{kraft:2001a}.

Cen~A was detected at MeV to GeV energies by all instruments on board the Compton Gamma-Ray Observatory (CGRO) in the period 1991 -- 1995, revealing a peak in the spectral energy distribution (SED) in $\nu F_\nu$ representation at  $\sim0.1$\,MeV with a maximum flux of about $\sim10^{-9}$\,erg\,cm$^{-2}$\,s$^{-1}$  \cite{steinle:1998a}. Recently, a detection of Cen~A at GeV energies has been reported by the \textit{Fermi} LAT team \cite{abdo:2009a:fermi:brightsourcelist:agn}. A tentative detection of Cen~A ($4.5\sigma$) at VHE during a giant X-ray outburst in the 1970's was reported by \cite{grindlay:1975a}. Subsequent VHE observations made with different instruments
resulted in upper limits.
Cen~A has been proposed as a possible source of ultra-high energy cosmic rays
(\cite{romero:1996a}, but see also \cite{lemoine:2008a}). 

\section{H.E.S.S. observations and results}

\begin{figure}
\centering
\includegraphics[width=0.45\textwidth]{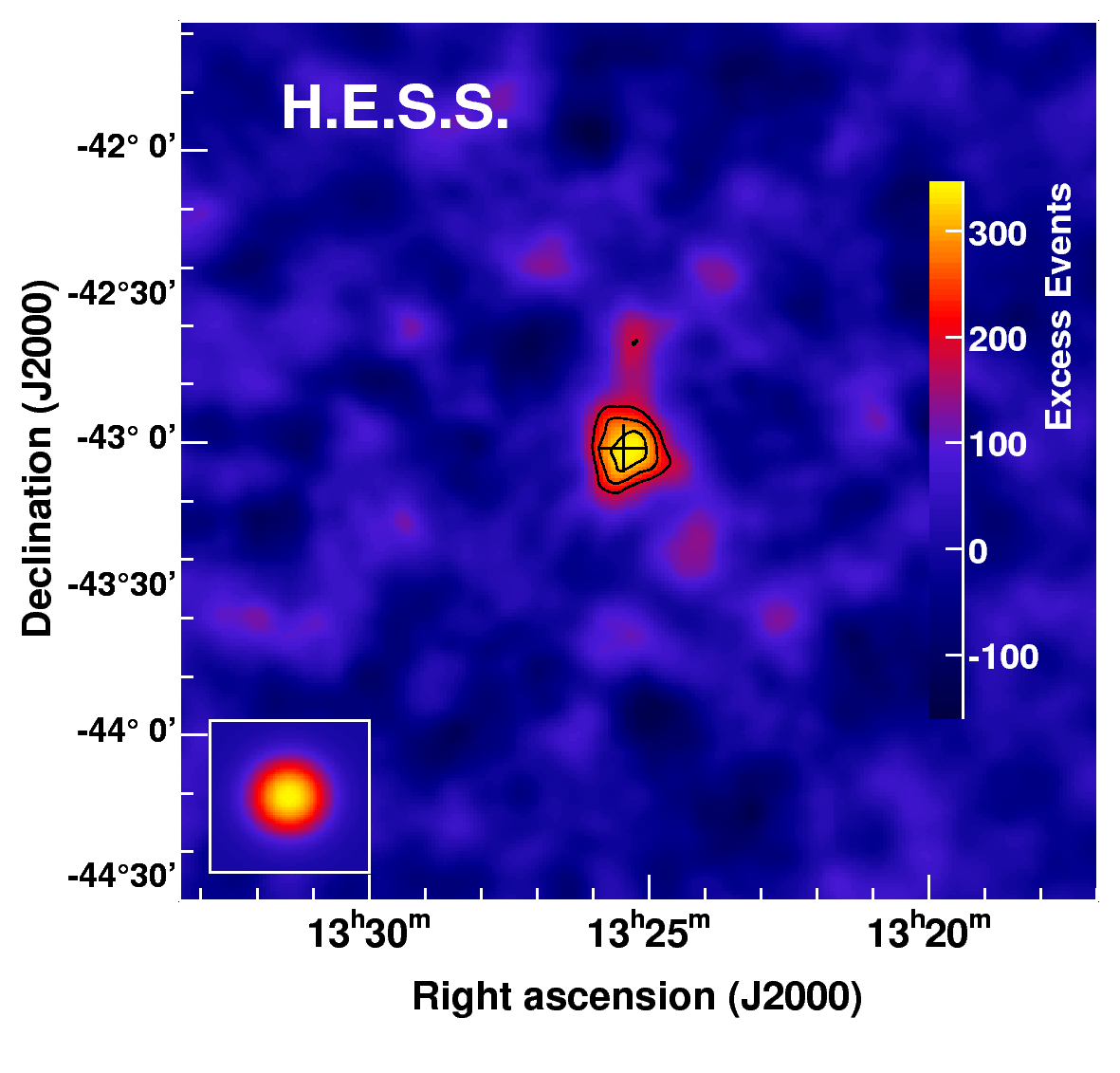} \hfill \includegraphics[width=0.45\textwidth]{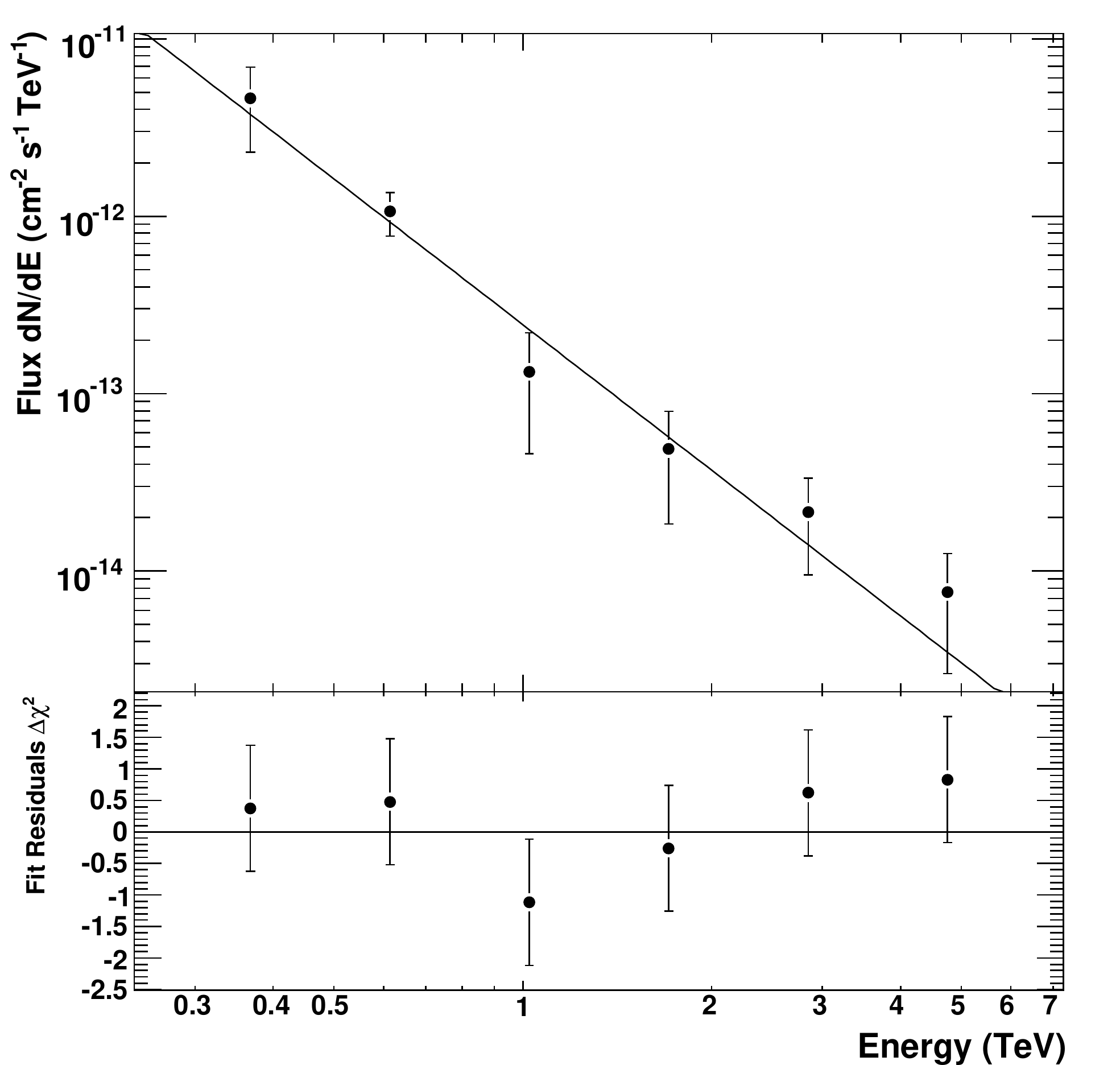}
\caption{\textit{Left:} Smoothed excess sky map of VHE $\gamma$-rays centered on the Cen~A radio core (cross) (contours: 3, 4, and 5$\sigma$). \textit{Right:} Differential energy spectrum of Cen~A as measured by H.E.S.S.}\label{Fig:1}
\end{figure}
 
The H.E.S.S. observations of Cen~A were performed between April  2004 and July 2008 (total live time: 115.0\,h; zenith angles: 20$^\circ$ to 60$^\circ$; mean zenith: $\sim24^\circ$). The data were analyzed with a standard Hillas-type analysis \cite{aharonian:2006:hess:crab} with an analysis energy threshold of $\sim250$\,GeV for a zenith angle of 20$^\circ$.

Figure \ref{Fig:1} shows the smoothed excess sky map of VHE $\gamma$-rays as measured with H.E.S.S. centered on the Cen~A radio core position. A clear excess at the position of Cen~A is visible. A point source analysis, using standard cuts as described in \cite{aharonian:2006:hess:crab}, was performed on the radio core position of Cen~A, resulting in the detection of an excess with a statistical significance of $5.0\,\sigma$. A fit of the instrumental point spread function to the uncorrelated sky map results in a good fit (chance probability $\sim0.7$) with a best fit position of  $\alpha_{J2000}=13^{\mathrm h}25^{\mathrm m}26.4^{\mathrm s}\pm4.6^{\mathrm s}_{\rm stat}\pm2.0^{\mathrm s}_{\rm syst}$, $\delta_{J2000}=-43^{\circ}0.7'\pm1.1'_{\rm stat}\pm30''_{\rm syst}$, well compatible with the radio core and the inner kpc jet region. Assuming a Gaussian surface-brightness profile, we derive an upper limit of 0.2$^\circ$ on the extension (95\% confidence level).

The measured differential photon spectrum (Fig.~\ref{Fig:1}) is well described by a power-law function $dN/dE = \Phi_0 \cdot (E/\mbox{1\,TeV})^{-\Gamma}$ with normalization $\Phi_0 = (2.45 \pm 0.52_{\mathrm{stat}} \pm  0.49_{\mathrm{sys}}) \times 10^{-13}$\,cm$^{-2}$\,s$^{-1}$\,TeV$^{-1}$ and photon index $\Gamma = 2.73 \pm 0.45_{\mathrm{stat}} \pm 0.2_{\mathrm{sys}}$. The integral flux above 250\,GeV, taken from the spectral fit, is $\Phi(E > 250\mathrm{\,GeV}) = (1.56 \pm 0.67_{\mathrm{stat}}) \times 10^{-12}$\,cm$^{-2}$\,s$^{-1}$, which corresponds to $\sim 0.8$\% Crab, or an apparent isotropic luminosity of  L($>$250\, GeV)$=2.6 \times 10^{39}$\,erg\,s$^{-1}$ (adopting a distance of 3.8\,Mpc).
No significant variability has been found on time-scales of 28\,min, nights and months (moon periods).
The results have been cross-checked with independent analysis and calibration chains and good agreement was found. More details can be found in \cite{aharonian:2009:hess:cena}.

\section{Discussion}
 
\begin{figure}
\centering
\includegraphics[width=0.6\textwidth]{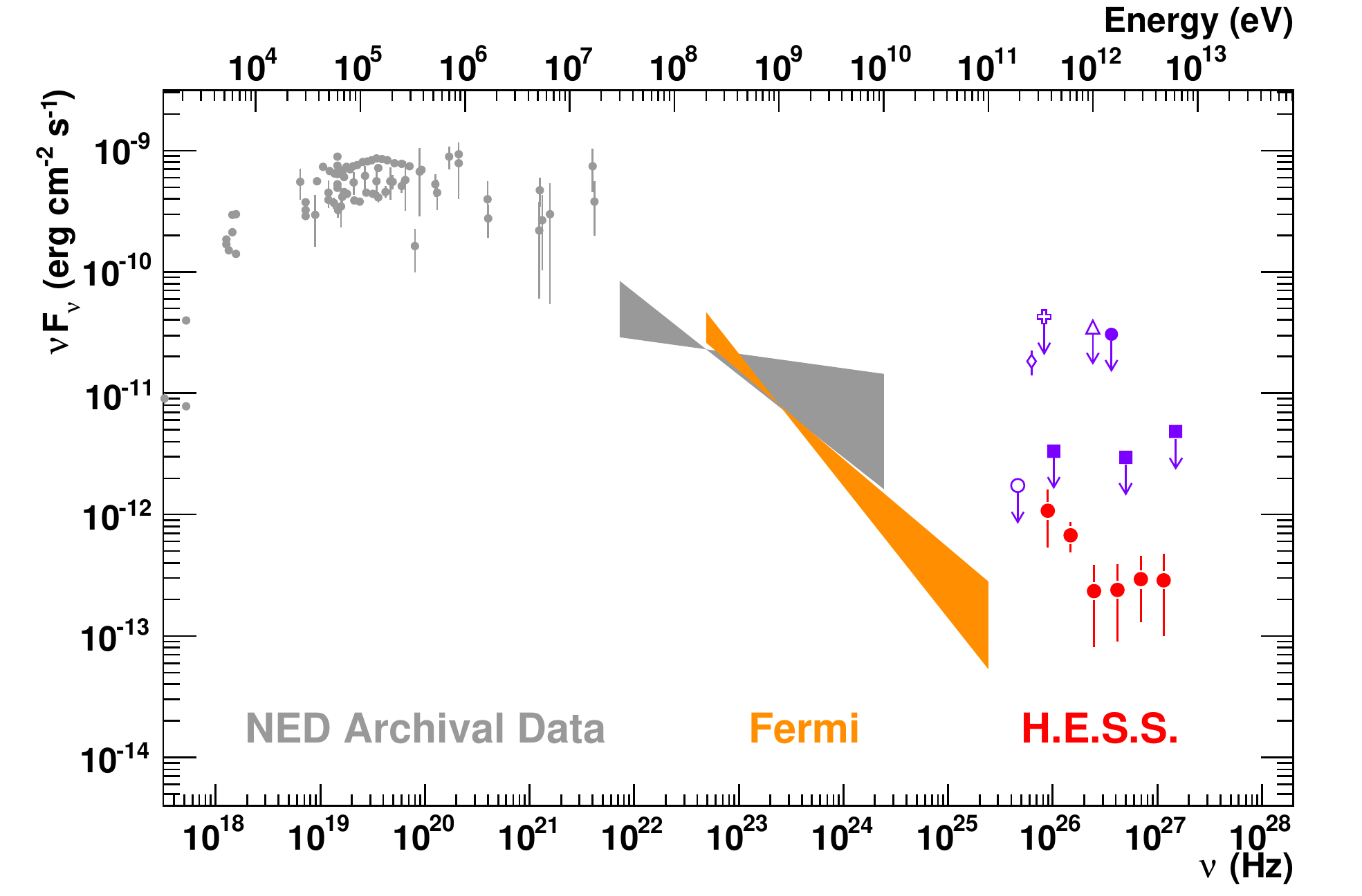}
\caption{High energy part of the spectral energy distribution of Cen~A.}
\label{Fig:SED}
\end{figure}
 
Figure~\ref{Fig:SED} shows the spectral energy distribution of Cen~A ranging from X-rays to the VHE regime.
The flux measured by H.E.S.S. is clearly below all previous upper limits in the VHE regime. Recently, the \textit{Fermi} LAT team reported a detection of Cen~A at GeV energies \cite{abdo:2009a:fermi:brightsourcelist:agn} (Fig.~\ref{Fig:SED} orange bow-tie). While a simple extrapolation of the reported power law function would result in a too low flux at $\sim$TeV energies, one will have to wait until the actual spectral points are released to conclude about the compatibility  of the H.E.S.S. and the \textit{Fermi} data. In addition, the data are not contemporaneous and variability cannot strongly be excluded from the H.E.S.S. data-set.

Several authors have predicted VHE emission from Cen~A, and more generally discussed VHE emission from radio galaxies.
A first class of models proposed the immediate vicinity of the supermassive black hole as the region of VHE emission, e.g. in pulsar-type scenarios \cite{neronov:2007a,rieger:2008a}.
A second class of models propose a mechanism similar to the mechanism at work in other VHE blazars \cite{bai:2001a,chiaberge:2001a}.
\cite{ghisellini:2005a} discussed a two-flow type model \cite[][]{sol:1989a,marcowith:1998a}, with a fast spine and a slower, mildly relativistic sheath propagating within the jet, which has been successfully applied to M\,87 \cite[][]{tavecchio:2008a}.
\cite{lenain:2008a} modeled the VHE emission of Cen~A with a multi-blob SSC model.
Extended VHE emission may also be expected from Cen~A.
In this context, \cite{stawarz:2006b} proposed that $\gamma$-rays emitted in the immediate vicinity of the active nucleus are partly absorbed by the starlight radiation in the host galaxy.
The created $e^\pm$ pairs are quickly isotropized and radiate VHE $\gamma$-rays by inverse Compton scattering the starlight radiation. The small size of the resulting isotropic pair halo ($\sim 4$ arcmin in diameter) is fully consistent with a point-like source for H.E.S.S., but could be resolved by the future CTA (Cherenkov Telescope Array)\footnote{http://www.cta-observatory.org/} observatory. 
Furthermore, hadronic models have been invoked to predict VHE emission from radio galaxies \cite{reimer:2004a}. 
The H.E.S.S. result do not yet strongly constrain these models.

Recently, \cite{croston:2009a} reported the detection of non-thermal X-ray synchrotron emission from the shock of the southwest inner radio lobe.
They investigated inverse Compton scattering the starlight radiation and the CMB from high energy particles in this lobe and predicted a VHE emission well compatible with the H.E.S.S. measurement. This study would suggest Cen\,A is analogous to a gigantic supernova remnant (SNR). While the position is $\sim3\sigma$ away from the best fit position of the VHE excess, it is well within the upper limit of the extension.

Cen\,A represents a rich potential for future VHE experiments. The current data are at the edge of differentiating the possible emitting regions. With higher sensitivity (factor 10), better astrometric accuracy and angular resolution (e.g. $\sim 5''$ and $\sim 1'$, respectively) \cite[][]{hermann:2007a}, CTA would allow the localization of the site of the VHE emission, and, possibly, reveal multiple VHE emitting sources within Cen~A. More generally, the detection of VHE emission from Cen~A together with the detection of M\,87 and the galactic center poses the question of whether VHE emission ($\gamma$-ray brightness) might be a general feature of AGN. While the sensitivity of current generation experiments is probably too low to answer this question, one can hope that the CTA experiment will be able to detect a large enough sample of sources to shed some light on this issue.


\section*{Acknowledgments}
{
\small
The support of the Namibian authorities and of the University of Namibia
in facilitating the construction and operation of H.E.S.S. is gratefully
acknowledged, as is the support by the German Ministry for Education and
Research (BMBF), the Max Planck Society, the French Ministry for Research,
the CNRS-IN2P3 and the Astroparticle Interdisciplinary Programme of the
CNRS, the U.K. Science and Technology Facilities Council (STFC),
the IPNP of the Charles University, the Polish Ministry of Science and 
Higher Education, the South African Department of
Science and Technology and National Research Foundation, and by the
University of Namibia. We appreciate the excellent work of the technical
support staff in Berlin, Durham, Hamburg, Heidelberg, Palaiseau, Paris,
Saclay, and in Namibia in the construction and operation of the
equipment.
This research has made use of the NASA/IPAC Extragalactic Database (NED) which is operated by the Jet Propulsion Laboratory, California Institute of Technology, under contract with the National Aeronautics and Space Administration, and NASA's Astrophysics Data System.
}

\def\Journal#1#2#3#4{{#4}, {#1}, {#2}, #3}
\def\NAT{Nature}
\def\AAA{A\&A}
\def\ApJ{ApJ}
\def\AJ{Astronom. Journal}
\def\Aph{Astropart. Phys.}
\def\ApJS{ApJSS}
\def\MNRAS{MNRAS}
\def\NIM{Nucl. Instrum. Methods}
\def\NIMA{Nucl. Instrum. Methods A}
\def\aj{AJ}%
\def\actaa{Acta Astron.}%
\def\araa{ARA\&A}%
\def\apj{ApJ}%
\def\apjl{ApJ}%
\def\apjs{ApJS}%
\def\ao{Appl.~Opt.}%
\def\apss{Ap\&SS}%
\def\aap{A\&A}%
\def\aapr{A\&A~Rev.}%
\def\aaps{A\&AS}%
\def\azh{AZh}%
\def\baas{BAAS}%
\def\bac{Bull. astr. Inst. Czechosl.}%
\def\caa{Chinese Astron. Astrophys.}%
\def\cjaa{Chinese J. Astron. Astrophys.}%
\def\icarus{Icarus}%
\def\jcap{J. Cosmology Astropart. Phys.}%
\def\jrasc{JRASC}%
\def\mnras{MNRAS}%
\def\memras{MmRAS}%
\def\na{New A}%
\def\nar{New A Rev.}%
\def\pasa{PASA}%
\def\pra{Phys.~Rev.~A}%
\def\prb{Phys.~Rev.~B}%
\def\prc{Phys.~Rev.~C}%
\def\prd{Phys.~Rev.~D}%
\def\pre{Phys.~Rev.~E}%
\def\prl{Phys.~Rev.~Lett.}%
\def\pasp{PASP}%
\def\pasj{PASJ}%
\def\qjras{QJRAS}%
\def\rmxaa{Rev. Mexicana Astron. Astrofis.}%
\def\skytel{S\&T}%
\def\solphys{Sol.~Phys.}%
\def\sovast{Soviet~Ast.}%
\def\ssr{Space~Sci.~Rev.}%
\def\zap{ZAp}%
\def\nat{Nature}%
\def\iaucirc{IAU~Circ.}%
\def\aplett{Astrophys.~Lett.}%
\def\apspr{Astrophys.~Space~Phys.~Res.}%
\def\bain{Bull.~Astron.~Inst.~Netherlands}%
\def\fcp{Fund.~Cosmic~Phys.}%
\def\gca{Geochim.~Cosmochim.~Acta}%
\def\grl{Geophys.~Res.~Lett.}%
\def\jcp{J.~Chem.~Phys.}%
\def\jgr{J.~Geophys.~Res.}%
\def\jqsrt{J.~Quant.~Spec.~Radiat.~Transf.}%
\def\memsai{Mem.~Soc.~Astron.~Italiana}%
\def\nphysa{Nucl.~Phys.~A}%
\def\physrep{Phys.~Rep.}%
\def\physscr{Phys.~Scr}%
\def\planss{Planet.~Space~Sci.}%
\def\procspie{Proc.~SPIE}%

\end{document}